\documentclass[revtex4]{emulateapj}
\usepackage{natbib}

\slugcomment{To Appear in ApJ}
\shorttitle{SN 2012aw Light Echo}
\shortauthors{Van Dyk et al.}

\begin{document}

\title{LEGUS Discovery of a Light Echo Around Supernova 2012aw}

\author{Schuyler D.~Van Dyk\altaffilmark{1}}
\author{Janice C.~Lee\altaffilmark{2}}
\author{Jay Anderson\altaffilmark{2}}
\author{Jennifer E.~Andrews\altaffilmark{3}}
\author{Daniela Calzetti\altaffilmark{4}}
\author{Stacey N.~Bright\altaffilmark{2}}
\author{Leonardo Ubeda\altaffilmark{2}}
\author{Linda J.~Smith\altaffilmark{5}}
\author{Elena Sabbi\altaffilmark{2}}
\author{Eva K.~Grebel\altaffilmark{6}}
\author{Artemio Herrero\altaffilmark{7,8}}
\author{Selma E.~de Mink\altaffilmark{9}}

\altaffiltext{1}{IPAC/Caltech, Mailcode 100-22,
  Pasadena, CA 91125, USA; email: vandyk@ipac.caltech.edu.}
\altaffiltext{2}{Space Telescope Science Institute, 3700 San Martin Drive, Baltimore, MD 21218, USA.}
\altaffiltext{3}{Steward Observatory, University of Arizona, Tucson, AZ 85721, USA.}
\altaffiltext{4}{Department of Astronomy, University of Massachusetts, Amherst, MA 01003, USA.}
\altaffiltext{5}{European Space Agency/Space Telescope Science Institute, Baltimore, MD 21218, USA.}
\altaffiltext{6}{Astronomisches Rechen-Institut, Zentrum f\"ur Astronomie
  der Universit\"at Heidelberg, M\"onchhofstr.\ 12-14,
  69120 Heidelberg, Germany.}
\altaffiltext{7}{Instituto de Astrof\'{i}sica de Canarias, V\'{i}a L\'{a}ctea s/n, E-38200 La Laguna (S.C.~Tenerife), Spain}
\altaffiltext{8}{Departamento de Astrof\'{i}sica, Universidad de La Laguna,
Avda.~Astrof\'{i}sico Francisco S\'anchez s/n, E-38071 La Laguna (S.C.~Tenerife), Spain}
\altaffiltext{9}{Anton Pannenkoek Astronomical Institute, University of Amsterdam, 1090 GE Amsterdam, The 
Netherlands}

\begin{abstract}
We have discovered a luminous light echo around the normal Type II-Plateau Supernova (SN)
2012aw in Messier 95 (M95; NGC 3351), 
detected in images obtained approximately two years after explosion with the Wide Field 
Channel 3 on-board the {\sl Hubble Space Telescope\/} ({\sl HST}) by the Legacy ExtraGalactic 
Ultraviolet Survey (LEGUS).  
The multi-band observations span from the near-ultraviolet through the optical (F275W,
F336W, F438W, F555W, and F814W).
The apparent brightness of the echo at the time was $\sim 21$--22 mag in all of these bands.
The echo appears circular, although less obviously as a ring, with an inhomogeneous surface brightness, in 
particular, a prominent enhanced brightness to the southeast.
The SN itself was still detectable, particularly in the redder bands.
We are able to model the light echo as the time-integrated SN light scattered off of diffuse interstellar dust 
in the SN environment. We have assumed that this dust is analogous to that in the Milky Way with 
$R_V=3.1$. The SN light curves that we consider also include models of the unobserved 
early burst of light from the SN shock breakout.
Our analysis of the echo suggests that the distance from the SN to the scattering dust elements along the
echo is $\approx$45 pc. The implied visual extinction for the echo-producing dust is consistent with estimates
made previously from the SN itself.
Finally, our estimate of the SN brightness in F814W is fainter than that measured for the red supergiant
star at the precise SN location in pre-SN images, possibly indicating that the star has vanished and 
confirming it as the likely SN progenitor.
\end{abstract}


\section{Introduction}

Light echoes from transient events provide a probe of both the circumstellar and interstellar structures
in the environment \citep[e.g.,][]{crotts91,xu95}, 
of the size distribution and chemical composition of the scattering dust 
\citep[e.g.,][]{sugerman03}, and of the detailed history of the outburst giving rise to the echo
\citep[e.g.,][]{rest12b}. Echoes also could provide a geometric distance to the event, independent of
the distance ladder, through analysis of polarized light \citep{sparks94,sparks96}.
A number of resolved light echoes have been discovered historically, including those 
of novae \citep{kapteyn01,ritchey01,sokoloski13},
unusual outbursting stars \citep{bond03}, erupting luminous blue variables \citep{rest12a,prieto14}, 
and ancient Galactic and Local Group supernovae \citep[SNe;][]{krause05,rest05,krause08,rest08a,rest08b}.
See \citet{rest12b} for a review.
Although light echoes from more recent extragalactic SNe could well be a common occurrence, 
observations which spatially resolve echoes are relatively rare, with only 11 occurrences known in total, 
as summarized in Table~\ref{tabhist}.
Inferences have been made for the existence of possible light echoes around a number of other SNe, 
mostly from reemission of SN light by dust in the infrared 
\citep{meikle06,welch07,mattila08,miller10,sugerman12}.
The main obstacle to resolving the echo 
is that the SN host galaxies must be relatively nearby, although, even then, 
the structures can only be revealed by the superior angular 
resolution of the {\sl Hubble Space Telescope\/} ({\it HST}).

The light echo of a SN results from the luminous ultraviolet (UV)/optical pulse scattered by dust in 
the SN environment. We essentially see a reflection of the event itself \citep{patat05}.
In particular, if the light echo is detected in the UV, we have a record of the bright, rapid flash of X-ray/UV
emission emerging as the SN shock breaks through the massive envelope around the progenitor star
\citep{crotts92}, an event that is elusive on account of its promptness and brevity 
\citep[e.g.,][]{quimby07,schawinski08,gezari08,gezari15}. 
Much of the circumstellar matter and small dust grains nearest to the SN will be
destroyed by the pulse, although more distant, larger grains will survive.
The observed light echo at a particular instant in time is the intersection of a dust sheet or filament with the 
(virtual) ellipsoid surface of constant arrival time, which results in nearly all cases as a circle or arc on the
sky.

\bibpunct[; ]{(}{)}{;}{a}{}{;}
We have discovered a resolved light echo around the normal SN II-Plateau (II-P) 
2012aw in the SB(r)b spiral host galaxy Messier 95 (M95; NGC~3351), during the course of
the Legacy ExtraGalactic UV Survey (LEGUS; GO-13364, PI: D.~Calzetti). LEGUS is a 
Cycle 21 {\it HST\/} Treasury Program which imaged 50 nearby ($\lesssim 12$ Mpc) galaxies in multiple 
bands with the Wide Field Camera 3 (WFC3) UVIS channel 
and the Advanced Camera for Surveys (ACS) Wide Field Channel. See \citet{calzetti15}
for a detailed introduction to the survey. LEGUS particularly concentrated
on obtaining images of these well-studied galaxies in the WFC3/UVIS F275W and F336W bands.

The SN has been extensively studied in the UV, optical, and near-infrared
by \citet{bayless13}, \citet{bose13}, and \citet{dallora14}.
\citet{yadav14} analyzed radio observations of the SN.
The SN progenitor was identified in archival {\sl HST\/} Wide-Field and Planetary Camera 2 (WFPC2)
F555W and F814W and ground-based $J$ and $K_s$ images
as a likely red supergiant (RSG) star by both \citet{vandyk12} and
\citet{fraser12}.
Both \citeauthor{fraser12}~and \citeauthor{vandyk12}~found that the RSG was cool 
($3500 \lesssim T_{\rm eff} ({\rm K}) \lesssim 4500$; although, \citeauthor{vandyk12}~more narrowly
limited $T_{\rm eff}$ to $\approx 3600$~K), must have had a dusty
circumstellar environment, and was therefore quite luminous ($L_{\star} \gtrsim 10^{5}\ L_{\odot}$) and of
relatively high initial mass\footnote{This actually should be referred to as a ``single-star-equivalent initial 
mass,'' since the majority of massive stars is expected to exchange mass or merge with a binary companion 
before explosion \citep{sana12}.} 
($M_{\rm ini} \simeq 15$--$25\ M_{\odot}$), based on the stellar evolutionary
models from \citet{eldridge08} and \citet{ekstrom12}, respectively.
\citet{kochanek12} 
reanalyzed the data from both papers, employing a different approach for the treatment of the circumstellar
extinction, 
and argued that the progenitor had to have been 
less luminous ($L_{\star} < 10^{5}\ L_{\odot}$) and less massive ($M_{\rm ini} < 15\ M_{\odot}$).
Van Dyk et al.~(in preparation) have revisited the progenitor photometry and conclude that the star
likely had $L_{\star} \approx 10^{4.9}\ L_{\odot}$ 
and  $M_{\rm ini} \approx 12\ M_{\odot}$.
\citet{jerkstrand14} modeled late-time spectra from the nebular phase and found that the emission lines
from likely nucleosynthetic products indicate a progenitor initial mass in the range 14--$18\ M_{\odot}$.

In this paper we describe the observations and analysis of the light echo and the ramifications for the
properties of the SN and its progenitor.
We assume a reddening- and metallicity-corrected distance modulus to the host galaxy
of $30.00 \pm 0.09$ mag \citep[measured using Cepheids;][]{freedman01}, 
which is a distance of $10.0 \pm 0.4$ Mpc.
The inclination of M95 is relatively low, at $41\arcdeg$, with position angle $192\arcdeg$ \citep{tamburro08},
so internal line-of-sight effects should be minimal.
The SN occurred $58{\farcs}3$ W and $115{\farcs}8$ S of the host galaxy nucleus, in a region of the galaxy
with no obvious signs of recent massive star formation. 
\citet{bose13} adopted 2012 March 16.1 (JD 2456002.6 $\pm$ 0.8) as the time of the SN explosion.
\citet{dallora14} adopted March 16.0 (JD 2456002.5 $\pm$ 0.8).
These are essentially the same for the purposes of our analysis below.
UT dates are used throughout.

\begin{deluxetable}{ccccc}
\tablenum{1}
\tablecolumns{5}
\tablewidth{0pt}
\tablecaption{Spatially Resolved Supernova Light Echoes\tablenotemark{a}\label{tabhist}}
\tablehead{
\colhead{SN} & \colhead{Type} & \colhead{Host} & \colhead{Distance} & \colhead{References} \\
\colhead{} & \colhead{} & \colhead{Galaxy} & \colhead{(Mpc)} & \colhead{} 
}
\startdata
1987A & II-P & LMC & 0.05 & 1,3,15 \\
1991T & Ia & NGC 4527 & 15 & 9 \\
1993J & IIb & M81 & 3.4 & 7,10 \\
1995E & Ia & NGC 2441 & 49.6 & 8 \\
1998bu & Ia & M96 & 9.9 & 2 \\
2003gd & II-P & M74 & 10.0 & 11,12 \\
2006X & Ia & M100 & 15.2 & 14 \\
2007af & Ia & NGC 5584 & 23 & 5 \\
2008bk & II-P & NGC 7793 & 3.4 & 13 \\
2012aw & II-P & M95 & 10.0 & 16 \\
2014J & Ia & M82 & 3.5 & 4 \\
\enddata
\tablenotetext{a}{Historical extragalactic SNe which have occurred since, and including, SN 1987A.}
\vspace{-0.3cm}
\tablerefs{
(1) \citet{bond90}; (2) \citet{cappellaro01}; (3) \citet{crotts88}; (4) \citet{crotts14}; (5) \citet{drozdov14}; 
(6) \citet{gouiffes88}; (7) \citet{liu03}; (8) \citet{quinn06}; (9) \citet{sparks99}; (10) \citet{sugerman02}; 
(11) \citet{sugerman05}; (12) \citet{vandyk06}; (13) \citet{vandyk13}; (14) \citet{wang08}; (15) \citet{xu94}; 
(16) This paper.
}
\end{deluxetable}

\begin{figure*}
\includegraphics[angle=0,scale=0.90]{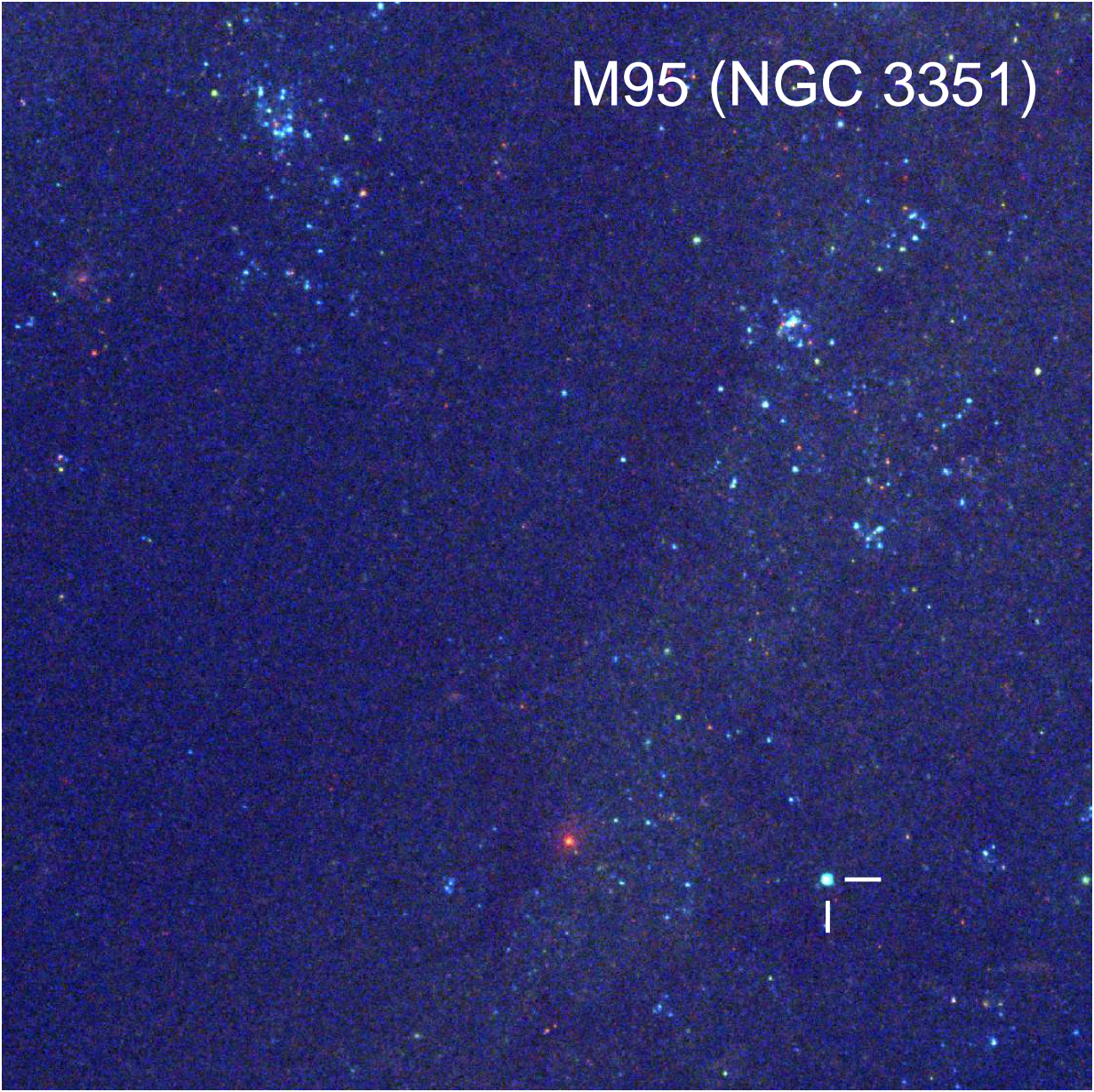}
\caption{A portion of the LEGUS {\sl HST\/} WFC3/UVIS image mosaics of M95 from 2014 April 23, 
as a three-color composite of bands F336W, F555W, and F814W.
The area shown is $40\arcsec \times 40\arcsec$.
A prominent light echo is seen at the site of the SN II-P 2012aw, as indicated by the tickmarks.
The SN occurred $58{\farcs}3$ W and $115{\farcs}8$ S of the galactic nucleus.
The SN age was $\sim 768.2$ d, or 2.10 yr, at the time.
North is up, and east is to the left.\label{figechocolor}}
\end{figure*}

\section{{\sl HST\/} Observations and Photometry}\label{observations}

M95 was initially observed by {\sl HST\/} for LEGUS with WFC3/UVIS on 2014 March 1, 
however, guide star tracking failed. The galaxy was reobserved on April 23 in bands F275W 
(total exposure time 2361 s), F336W (1062 s), F438W (908 s), F555W (1062 s), and F814W (908 s).
The pointing for the observations was designed intentionally to include the site of SN 2012aw in the total
field-of-view.
The imaging data were initially run through the standard STScI pipeline procedures for calibration.
However, as described in \citet{calzetti15}, 
further processing of the `flt' images included a pixel-based correction for the 
charge transfer efficiency (CTE) losses\footnote{Anderson, J., 2013, 
http://www.stsci.edu/hst/wfc3/tools/cte{\textunderscore}tools}. 
The corrected `flc' images were then combined
using the DrizzlePac software \citep{fruchter10,gonzaga12} to produce a distortion-corrected mosaic in 
each band. 

The midpoint of the WFC3 observations was at 2014 April 23 5:28:06, or JD 2456770.7. 
This is at a SN age of $\approx 768$ d, or 2.10 yr.

\begin{figure*}
\includegraphics[angle=0,scale=0.95]{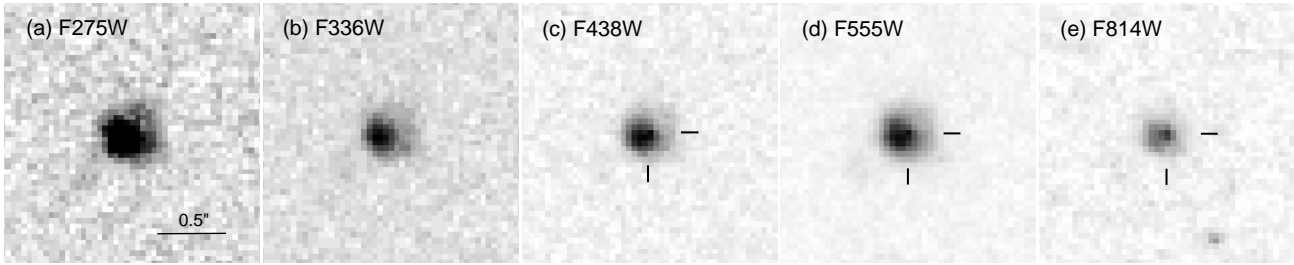}
\caption{A portion of the LEGUS {\sl HST\/} WFC3/UVIS image mosaics, showing the light echo in 
more detail, in bands ({\it a}) F275W, ({\it b}) F336W, ({\it c}) F438W, ({\it d}) F555W,
and ({\it e}) F814W. The greyscale has been set to the square-root of the image counts, to enhance the 
contrast.
The SN itself can still be seen, especially in the F438W,
F555W, and F814W images, as indicated by the tickmarks.
North is up, and east is to the left.\label{figecho}}
\end{figure*}

A three-color-composite image of the light echo is shown in Figure~\ref{figechocolor}.
The echo is conspicuously blue, which is not surprising, since we expect dust to scatter preferentially
the blue light from the SN.
Individual images of the echo are shown for all five bands in Figure~\ref{figecho}. 
The appearance of the echo is more of a circular patch, or blob, and less obviously a single ring.
If the echo were a thin ring, though, its profile would be convolved with the WFC3/UVIS PSF and 
would not be sharply defined. 
It is also entirely possible that we are actually seeing multiple rings, such as was observed for the interstellar
dust-scattered echo around SN 1987A \citep[][see his Figure 1]{crotts88}, just not as well resolved.
We know that the echo is a recent, post-SN apparition, since no evidence of it exists in the pre-explosion
{\sl HST\/} F555W and F814W images (see \citealt{vandyk12}, their Figure 1, and \citealt{fraser12}, their
Figure 2).
Using 20 fiducial stars in common between the pre-SN F814W image
and the LEGUS WFC3 F814W image in the immediate SN environment, we
can precisely associate the position of the SN progenitor star with the bright spot seen within the echo in 
Figure~\ref{figecho}(e), with a rms uncertainty of $0{\farcs}006$ (in right ascension) and $0{\farcs}009$
(in declination).
There is little question that this is the SN itself, still prominently seen in F814W, relative to the fainter echo
brightness.
The SN is also apparent in F438W and F555W, but less so in F275W and F336W.
A surface brightness enhancement to the southeast along the echo tends to dominate the echo's
appearance from F555W blueward.
This implies that the scattering dust is not uniformly distributed, 
which is consistent with the filamentary nature of interstellar dust 
and its subsequent effect on the structure of the observed light echoes \citep[e.g.,][]{rest11a}. 

We performed aperture photometry within IRAF\footnote{IRAF, the Image Reduction and Analysis
  Facility is distributed by the National Optical Astronomy
  Observatory, which is operated by the Association of Universities
  for Research in Astronomy (AURA) under cooperative agreement with
  the National Science Foundation (NSF).}
of the entire light echo on the drizzled image mosaics
in each band,
through a range of apertures with increasing radii, in order to estimate the uncertainty in the measurement. 
A sky annulus was used to estimate the background, which was relatively low in each band.
The VEGAMAG zero points for an infinite aperture for the WFC3 
bandpasses\footnote{http://www.stsci.edu/hst/wfc3/phot{\textunderscore}zp{\textunderscore}lbn.}
were adopted.
We list our photometry for the echo in Table~2. The flux of the echo, $F_{\rm echo}$, 
which is also listed in the table for 
each band, was established using these WFC3 zeropoints.

We emphasize that the photometry of the echo in Table~2 actually 
includes both the echo {\em and\/} the SN, particularly at the redder bands. 
To fully analyze the properties of the echo we needed to remove the SN flux 
from the total.  This was the most complicated step in our analysis and 
involved significant uncertainty.  We started with the three F814W exposures,
since the SN was brightest relative to the echo in that band.  Even in
F814W, though, the emission surrounding the SN is not necessarily spatially 
flat, so defining a sky background around the point source is nearly 
impossible.  We searched for the location and flux of the SN point source 
with the understanding that the SN should be the sharpest feature in the 
image and the nebulous background should be relatively smooth.  To this 
end, we extracted a 19$\times$19-pixel raster about the target in each 
flc exposure.  We then explored an array of positions and fluxes for the 
point source.  For each trial position and flux, we subtracted the 
corresponding point-spread function (PSF) from the raw exposures using a library PSF tailored for 
the particular filter at the particular location in the detector.  These 
library PSFs were constructed from a well-dithered set of data taken of 
the globular cluster Omega Centauri soon after WFC3 was installed on-board 
{\sl HST}; they have proven to be good to better than $\sim$2\%.  

We then examined the residuals and identified the optimal location and 
flux for the SN to be the one that left the post-subtraction residuals 
as smooth as possible.  To evaluate the smoothness of each residual image, 
we compared that image with a smoothed version of itself. The smoothed 
version is obtained by convolving the image with a 5$\times$5 quadratic 
smoothing kernel, which is equivalent to fitting for a six-parameter, 
two-dimensional quadratic centered on each pixel.  In essence, this operation 
reduces the number of degrees-of-freedom by a factor of 25/6.  The image 
that was closest to its smoothed version was deemed to be the smoothest.  
The optimal location and flux of the point source was deemed to be the 
one that left the image as close as possible to the smooth version of itself.

Using the above strategy, we determined a best-fit location and flux for
the SN in each of the F814W flc exposures.  When the positions were transformed
into the reference frame, they agreed to within a rms uncertainty of 0.09 pixel in $x$
and 0.03 pixel in $y$.  Using this best-fit position, we then found the
flux of the point source in each F814W exposure that made the resulting
image as smooth as possible and arrived at a flux of 650$\pm$75 e- for 
an exposure time of 359 s.  The PSFs were normalized to correspond to the 
flux within a 10-pixel aperture, so that is the effective aperture for these
measured fluxes.  We applied a similar approach to determine a best-fit 
flux for the SN in each of the other filters (850$\pm$150 e- in F555W, 
400$\pm$45 e- in F438W, 50$\pm$75 e- in F336W, and 100$\pm$50 e- in F275W), 
using the master-frame position fit from the F814W exposures. 

We used the same PSFs to measure fluxes for a number of other isolated 
point sources in the images, to tie this photometry to the Vega magnitudes for
stars in the images produced by Dolphot \citep{dolphin00}.  The SN flux was then similarly
scaled in each band; the resulting magnitudes are listed in 
Table~2. The uncertainties for the SN magnitudes in each band 
also include (via quadrature sums) the uncertainties in the scaling, 
although these are relatively small, compared to the uncertainties due 
to the smoothed model fitting. Finally, we list in Table~2
the resulting flux, $F^{\rm corr}_{\rm echo}$, obtained from subtracting 
the SN flux from our aperture measurements of the echo flux.
The fractional contribution of the SN light to the total echo flux is $\sim 2$\% in F275W,
$\sim 1$\% in F336W, $\sim 5$\% in F438W, $\sim 9$\% in F555W, and $\sim 24$\% at F814W. 

\begin{figure}
\epsscale{0.8}
\plotone{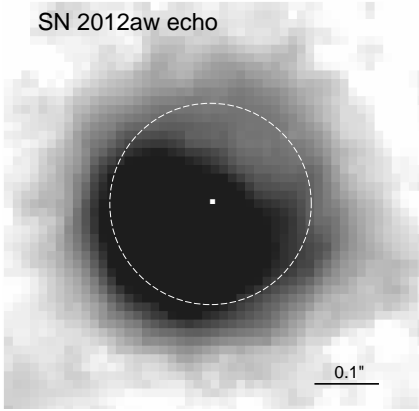}
\caption{A close-up of the SN 2012aw light echo in the sum of images in F275W, F438W, and F555W,
after the SN flux had been removed. The images were produced by $3\times$-supersampling and coadding
the individual exposures in each band. The greyscale has been set to the square-root of the image counts, 
to enhance the contrast. The echo could consist of a ring, or partial arcs, although this is not entirely evident,
e.g., there is the surface brightness enhancement seen toward the southeast and no central void indicating
a lack of dust in the plane of the SN. Nonetheless, the white circle represents our estimate of a
3.9-pixel ring radius, with a conservative uncertainty of $\pm 0.5$ pixel (see Figure~\ref{figcuts}). 
The white dot indicates the position of the now-subtracted SN within the echo.
At the assumed distance of M95, this corresponds to a radius of $7.6 \pm 1.0$ pc.\label{figsuper}}
\end{figure}

Furthermore, we had $3\times$-supersampled and coadded the individual flc exposures in each band,
with the SN flux removed, to produce images of just the light echo. In Figure~\ref{figsuper} we show
the F275W, F438W, and F555W images summed together, in order to increase the overall 
signal-to-noise ratio of the fainter features of the echo. We did this primarily to reveal any ring-like
structure. As one can see, the shape of the echo is still quite patchy and irregular, with the southeast
enhancement still dominating the emission, although hints of a ring, or partial arcs, are possibly visible.

\begin{deluxetable*}{ccccc}
\tablenum{2}
\tablecolumns{5}
\tablewidth{0pt}
\tablecaption{Photometry of SN 2012aw and Its Light Echo\tablenotemark{a}\label{tabecho}}
\tablehead{
\colhead{Band} & \colhead{$m_{\rm echo}$} & \colhead{$F_{\rm echo}$} &
\colhead{$m_{\rm SN}$} & \colhead{$F^{\rm corr}_{\rm echo}$} \\
\colhead{} & \colhead{(mag)} & \colhead{(erg cm$^{-2}$ s$^{-1}$ \AA$^{-1}$)} &
\colhead{(mag)} & \colhead{(erg cm$^{-2}$ s$^{-1}$ \AA$^{-1}$)} 
}
\startdata
F275W & $21.08 \pm 0.02$ & $1.39 {\pm 0.03} \times 10^{-17}$ & $25.35 \pm 0.61$ & 
$1.35 {\pm 0.04} \times 10^{-17}$ \\
F336W & $21.28 \pm 0.02$ & $1.00 {\pm 0.02} \times 10^{-17}$ & $26.12 \pm 1.00$ & 
$9.88 {\pm 0.03} \times 10^{-18}$ \\
F438W & $22.00 \pm 0.02$ & $1.06 {\pm 0.02} \times 10^{-17}$ & $25.29 \pm 0.14$ & 
$1.01 {\pm 0.03} \times 10^{-17}$ \\
F555W & $21.79 \pm 0.01$ & $7.64 {\pm 0.05} \times 10^{-18}$ & $24.38 \pm 0.20$ & 
$6.94 {\pm 0.02} \times 10^{-18}$ \\
F814W & $22.26 \pm 0.02$ & $1.44 {\pm 0.03} \times 10^{-18}$ & $23.79 \pm 0.13$ &  
$1.09 {\pm 0.01} \times 10^{-18}$ \\
\enddata
\tablenotetext{a}{We adopted for both $F_{\rm echo}$
and $F^{\rm corr}_{\rm echo}$ the WFC3 VEGAMAG flux zero points at
http://www.stsci.edu/hst/wfc3/phot{\textunderscore}zp{\textunderscore}lbn.}
\end{deluxetable*}

\section{Analysis of the Echo}\label{analysis}

\begin{figure}
\plotone{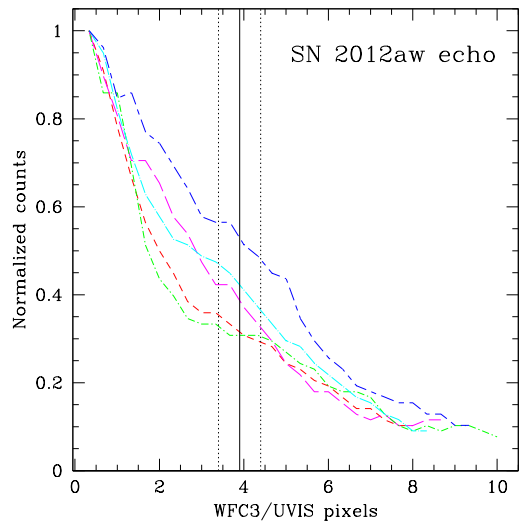}
\caption{Profile cuts through the SN 2012aw light echo, shown in Figure ~\ref{figsuper}, starting
at the echo center and progressing toward the north (short-dashed line), northeast (long-dashed line), 
northwest (dotted short-dashed line), west (dotted long-dashed line), and southwest (short-dashed long-dashed line),
intentionally avoiding the bright surface brightness enhancement to the southeast.
We estimated from these cuts that the nominal radius of the echo is 3.9 WFC3/UVIS pixels (solid line), with 
an uncertainty of $\pm 0.5$ pixel (dotted line).\label{figcuts}}
\end{figure}

As pointed out, it is not clear from Figure~\ref{figsuper} that what we are seeing is a well-defined ring, with an
expected central void indicating a lack of dust in the plane of the SN. 
We therefore cannot exclude at this point that the SN is actually immersed in a diffuse dust cloud.
Nonetheless, we determined the radius of any ring-like structure by taking cross-cut flux profiles 
through the echo from its center, avoiding the bright surface brightness enhancement. 
See Figure~\ref{figcuts}. We estimated the radius 
to be 3.9 pixels, with a conservative uncertainty of $\pm 0.5$ pixel. 
At the UVIS plate scale, this radius corresponds to
$\theta=0{\farcs}16 \pm 0{\farcs}02$. We show our estimate of a ring in Figure~\ref{figsuper}.
We can approximate the light echo ellipsoid as a paraboloid, since the SN's distance from the 
Earth, $D$, is so much larger than the geometric dimensions of the echo 
\citep[e.g.,][]{che86,schaefer87}.  
The perpendicular linear distance of the line-of-sight to the SN from the line-of-sight to the echo 
is $b=D \theta$, where $\theta$ is the angular distance between the two lines-of-sight.  
At the distance $D$ to M95, $b=7.6 \pm 1.0$ pc.   

The distance from the SN to any scattering dust element along the echo,
$r=l+ct$, can be obtained from $r^2 = b^2 + l^2$, where $l$ is the distance
from the SN to the echo along the line-of-sight \citep{couderc39}.  For $ct=0.64$ pc (2.10 ly), we
find that $l= 44.1$ pc and $r=44.7$ pc, with an uncertainty of $\pm 15.0$ pc, resulting from  
the uncertainties in the estimates of both $\theta$ and $D$.
The echo is therefore most likely due to interstellar dust. 
We note that much or all of any pre-existing circumstellar  
dust was likely destroyed by the SN X-ray/UV flash \citep{vandyk12}.

In modeling the echo, even with the uncertainties in the actual distribution of the diffuse interstellar dust, we 
have assumed that the echo arises from single scattering in
a thin dust sheet located between us and the SN, and that the sheet thickness is
much smaller than the distance between the sheet and the SN. 
The scattered flux $F$ at time $t$ from the echo at a given wavelength or bandpass is then 

\begin{equation}
\label{echomodel}
F^{\rm corr}_{\rm echo} (t) = \int_0^t F_{\rm SN} (t-t^{\prime}) f(t^{\prime}) dt^{\prime},
\end{equation}

\noindent where $F_{\rm SN} (t-t^{\prime})$ is the fluence of the SN at time
$t-t^{\prime}$, and $f(t^{\prime})$ is the impulse response function, i.e., the fraction of
light scattered by the echo toward us, in units of inverse time, which depends on the nature of the dust  
and the echo geometry \citep{che86,cappellaro01,patat05}.  
The total SN light is treated
as a short pulse over which the SN flux is constant, i.e., 
$F_{\rm SN} {\Delta}t_{\rm SN} = \int_0^{\infty} F_{\rm SN}(t) dt$ \citep{cappellaro01,patat05}.
The SN fluence is the integral of the light curves  in each band with respect to time.  

\begin{figure}
\plotone{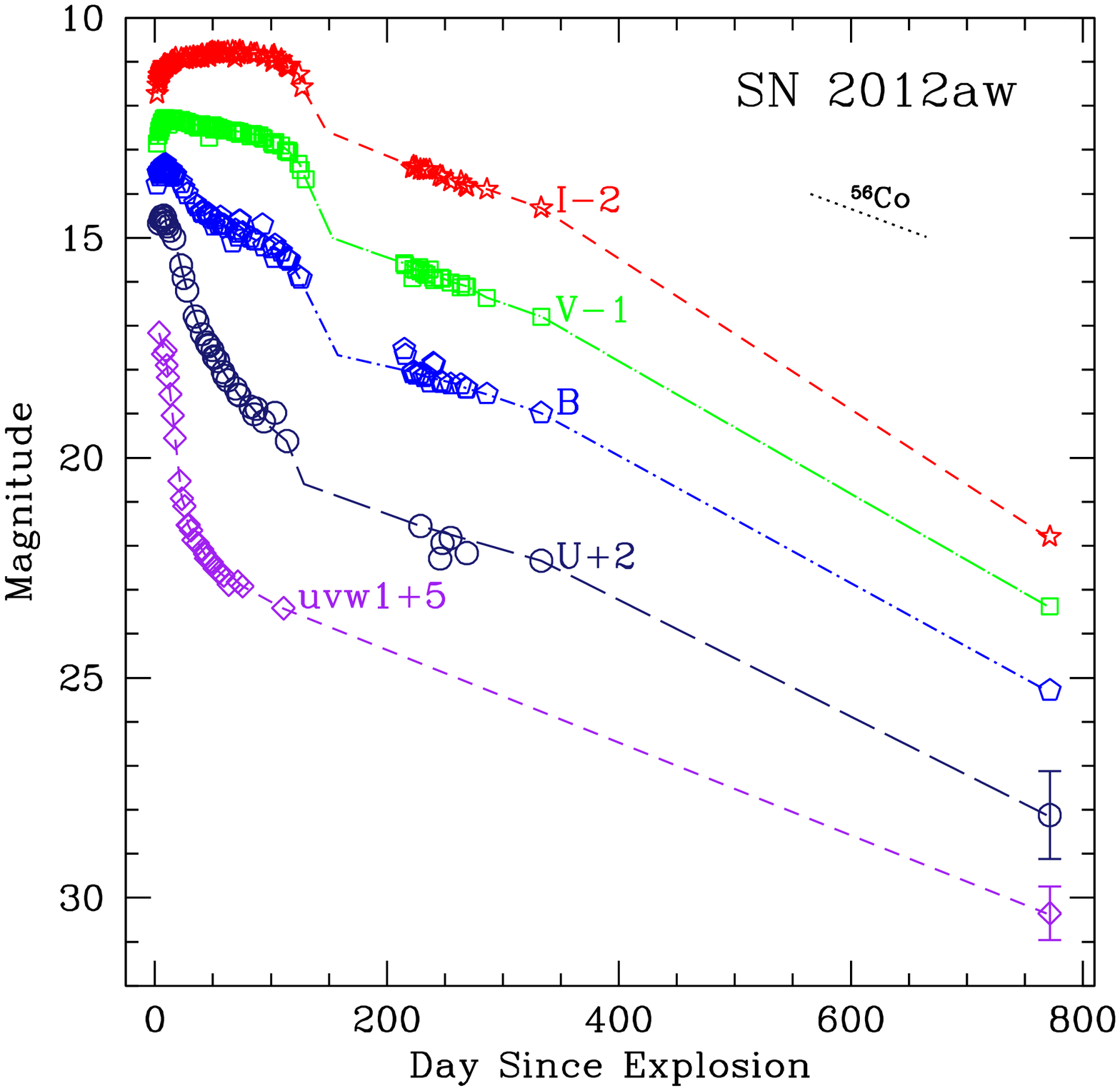}
\caption{Observed light curves (open points) for SN 2012aw from \citet{bayless13}, \citet{bose13}
and \citet{dallora14} at {\it SWIFT\/} $uvw1$ (diamonds), $U$ (circles), $B$ (pentagons), $V$ (squares), and
$I$ (five-pointed stars), which also include our estimate in each band of the SN brightness on day $\sim$768. 
The lines are interpolations of the observed data points in each band, although we have inserted our best
estimate of the plateau end and commencement of the exponential tail to the light curves.
Also shown for comparison is the expected decline rate as a result of $^{56}$Co decay.\label{latelc}}
\end{figure}

The light curves that we considered here are the combination of those presented in \citet{bose13}
and \citet{dallora14} in $UBVI$, which approximately match what we see from the echo in F336W,
F438W, F555W, and F814W, and the  {\it SWIFT\/} $uvw1$ data from \citet{bayless13},
where $uvw1$  is the most similar {\it SWIFT\/} bandpass to F275W. See Figure~\ref{latelc}.
Monitoring in $BVI$ began when the SN was at age $\approx$ 2 days, while the $U$ and $uvw1$
data commenced at $\approx$ 4 days.
Of course, at the very earliest ages the SN light is dominated by the shock breakout, which is
particularly important in the UV as a luminous flash. However, the shock breakout was mostly not
observed in the case of SN 2012aw. So, we have simulated what might have occurred, via shock breakout
models, specifically, the analytical expressions for RSGs from \citet{nakar10} and the hydrodynamical model
for a $13\ M_{\odot}$ RSG from \citet{tominaga11}. Ehud Nakar graciously provided his algorithms, which we
ran assuming an initial mass $M_{\rm ini}=12\ M_{\odot}$, a 
radius\footnote{Van Dyk et al., in preparation, have also revised the estimated radius of the SN
2012aw progenitor to be $R=750\ R_{\odot}$.} $R\approx 750\ R_{\odot}$, and an 
explosion energy of $10^{51}$ erg, at the central wavelengths of the five bandpasses. 
The coefficients in the \citet{nakar10} model have since been calibrated through numerical simulations of the
explosion of an analytical star (T.~Shussman et al., in preparation), and we used these updates.
Nozomu Tominaga also provided us with the output from his RSG model through the five
filter bandpasses. We have adjusted this model to $R=750\ R_{\odot}$ and converted from AB magnitudes
to Vega magnitudes.

\begin{figure}
\epsscale{0.90}
\plotone{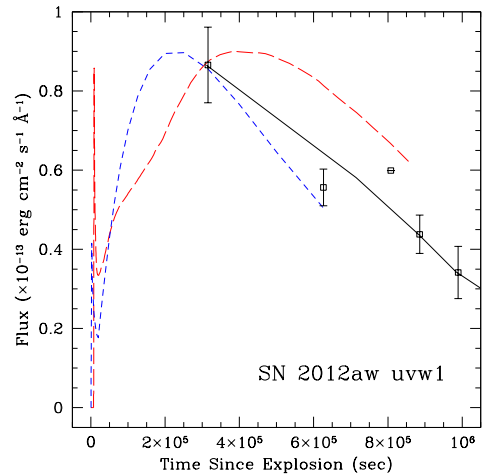}
\caption{Comparison of the {\it SWIFT\/} $uvw1$ light curve \citep{bayless13} with the 
adopted shock-breakout flash models in the UV within the first $8.6 \times 10^5$ s ($\sim 10$ d).  
The {\it SWIFT\/} data (open squares, solid
line) have been reddening-corrected, assuming a total $A_V=0.24$ mag and the \citet{cardelli89} reddening
law. The short-dashed line is the model light curve using
the analytical formulation from \citet{nakar10}. The long-dashed line is the model light curve from 
\citet{tominaga11}. See the text.\label{earlylc}}
\end{figure}

We show, specifically, the very early-time observed and model UV light curves in Figure~\ref{earlylc}. 
Here we are showing the curves in flux units, after reddening correction for an assumed total extinction,
$A_V=0.24$ mag (see below), assuming the \citet{cardelli89} reddening law with $R_V=3.1$.
The behavior of the two models is relatively similar, although they exhibit notable differences during the 
first $\sim 8$ d after explosion. 
The post-explosion peak of the breakout in the UV is noticeably more pronounced in the \citet{tominaga11} 
model than in the \citet{nakar10} model. The secondary peak at $\sim 5$ d in the 
\citeauthor{tominaga11}~model is as 
luminous, if not somewhat more so, as the initial peak near $\sim 0$ d. 
The secondary peak emission for the \citeauthor{nakar10} model occurs earlier (at $\sim 2.5$ d), and is more
in line with the first observed datapoint, than for the \citeauthor{tominaga11}~model.
These secondary peaks and the declines 
thereafter undershoot (for \citeauthor{nakar10}) and overshoot (for \citeauthor{tominaga11})
the observed UV light curve. We therefore truncate both model UV
curves at $\sim 4$ d.
The behavior of the model light curves, relative to the observations, is quite similar in the other bands, 
although the initial post-explosion flash diminishes in brightness toward the longer wavelengths.
We merely needed to truncate the models
before the first datapoint, as we did in the UV, to match the observations in the other bands.
The fact that both sets of shock breakout models tend to agree quite well in flux with the
early observed data, requiring very little adjustment, implies that the input parameters for those models, i.e., the 
progenitor initial mass and SN explosion energy, are reasonable representations of SN 2012aw and its progenitor star.

\begin{deluxetable*}{cccc}
\tablenum{3}
\tablecolumns{4}
\tablewidth{0pt}
\tablecaption{Time-Integrated SN 2012aw Light Curves\tablenotemark{a}\label{tabfluence}}
\tablehead{
\colhead{Band} & \colhead{Fluence [no flash]} & \colhead{Fluence [+Nakar model]} &
\colhead{Fluence [+Tominaga model]} \\
\colhead{} & \colhead{(erg cm$^{-2}$ \AA$^{-1}$)} & \colhead{(erg cm$^{-2}$ \AA$^{-1}$)} &
\colhead{(erg cm$^{-2}$ \AA$^{-1}$)} 
}
\startdata
$uvw1$ & $6.79 \times 10^{-8}$ & $9.03 \times 10^{-8}$ & $8.70 \times 10^{-8}$ \\
$U$ & $1.02 \times 10^{-7}$ & $1.12 \times 10^{-7}$ & $1.12 \times 10^{-7}$ \\
$B$ & $1.71 \times 10^{-7}$ & $1.72 \times 10^{-7}$ & $1.73 \times 10^{-7}$ \\
$V$ & $2.06 \times 10^{-7}$ & $2.07 \times 10^{-7}$ & $2.08 \times 10^{-7}$ \\
$I$ & $1.31 \times 10^{-7}$ & $1.31 \times 10^{-7}$ & $1.31 \times 10^{-7}$ \\
\enddata
\tablenotetext{a}{The conversion from {\sl SWIFT\/} and Johnson-Cousins Vega magnitudes 
to flux assumes the following zero points 
for Vega, as synthesized from the {\sl HST}/STIS calibration 
spectrum of the star using the package STSDAS/Synphot : 
$3.99 \times 10^{-9}$ ($uvw1$), 
$4.21 \times 10^{-9}$ ($U$), 
$6.30 \times 10^{-9}$ ($B$), 
$3.61 \times 10^{-9}$ ($V$), and
$1.19 \times 10^{-9}$ ($I$) erg cm$^{-2}$ s$^{-1}$ \AA$^{-1}$.}
\end{deluxetable*}

We have also included in the observed light curves our estimates of the SN 
brightness in the 2014 images, as shown in Figure~\ref{latelc}, where we assume F275W $\sim uvw1$, 
F336W $\sim U$, F438W $\sim B$, F555W $\sim V$, and F814W $\sim I$. All of the light curves at late
times declined in a manner roughly consistent with radioactive decay from day $\sim 340$ to day $\sim 768$, 
although the $I$ curve appears to have declined more steeply, while the $uvw1$ curve somewhat less
steeply.

After correction for the  total extinction to the SN (see below), 
we integrated, first, just the observed light curves in each band (neglecting the comparatively small 
uncertainties in the observed SN  photometry) and, then, these light curves including the two
early-time flash models. Note that for the observed light curves shown in Figure~\ref{latelc} we are missing 
the end of the plateau in the redder bands. (In the UV no indication 
exists of a pronounced plateau, so this is not an issue for this band.) 
For this reason, in particular, we interpolated the light
curves in each band from the observed datapoints and estimated the approximate end of the plateau 
and commencement
of the exponential decay phase of the light curves, based on the slope of the decline inferred from the initial 
data points, post-plateau, and the slope of the data points on the decline after day $\sim 200$ in each band. 
The slopes of the decline in the light curves after the end of the observed data 
(at day $\sim 340$ in $UBVI$, day $\sim 110$ in $uvw1$) were interpolated up to our estimates on day 
$\sim 768$.
Our integrations, then, were over these interpolated light curves, rather than the actual observed 
data. However, we consider the uncertainties in the total fluence introduced by this assumption to be small, 
since the behavior of our interpolated light curves are quite similar to that of the observed light curves of the 
prototypical, normal SN II-P 1999em, for which the end of the plateau was observed; see \citet{hamuy01} and 
\citet{leonard02}.
We adopted Vega as the flux zero point, obtained via synthetic photometry of Vega's observed {\sl HST\/} 
Space Telescope Imaging Spectrograph (STIS) spectrum using 
STSDAS/SYNPHOT\footnote{We obtained the SWIFT $uvw1$ filter transmission function from 
http://svo2.cab.inta-csic.es/theory/fps3/.} in IRAF. We list the results in Table~3.
We see that the flash models in the UV increase the overall fluence by $\sim 22$\% 
(\citeauthor{tominaga11}~model) to $\sim 25$\% (\citeauthor{nakar10}~model). This is in better
agreement with the observations, since  
the flux from the echo is relatively high in the UV. The early-time shock-breakout models contribute 
$\sim 9$\% additional flux at $U$, but only negligible additional amounts at $BVI$.

Treating $F_{\rm SN}$ as the SN maximum flux in each band, we find that the SN pulse
duration is 3.8, 21.4, 51.6, 110.3, and 141.2 days in $uvw1$, $U$, $B$, $V$, and $I$, respectively.
The maximum value of $w$, the effective pulse width \citep{sugerman02,sugerman03},  
is then $\approx 0.7$ pc (from $I$).
The observed echo thickness, $\Delta b$, is unresolved and is conservatively less than the FWHM of a 
single stellar profile, $\lesssim$2.0 WFC3 pixels, or, $\Delta b \lesssim 3.9$ pc.
It follows from \citet{sugerman03} that the dust thickness, $\Delta l$, is $\lesssim 46$ pc.
This value is clearly not less than our estimate of the dust sheet distance, $l$. 
However, since we can only place an upper limit on $\Delta l$ (given the upper limit on $\Delta b$), 
we infer that its actual value is likely far smaller. 
Our assumptions for the 
dust scattering, above, therefore likely still hold.

Following, e.g., \citet{cappellaro01} and \citet{patat05}

\begin{equation}
\label{eqresponse}
f(t)= {{c N_H} \over r}{\Phi({\theta})} C_{\rm sca} \int \phi(a) da,
\end{equation}

\noindent 
where $c$ is the speed of light, $N_H$ is the H column density, $\Phi({\theta})$ is
the scattering ``phase function,'' $C_{\rm sca}$ is the scattering cross section, and 
$\phi(a)$ is the grain size distribution for grain radius $a$.
The classical \citet{hen41} $\Phi({\theta})$ is not generally applicable in this case. As \citet{dra03}
points out, the phase function in the UV is strongly forward-scattering, particularly as the
scattering angle becomes small ($\theta \lesssim 10\arcdeg$). \citeauthor{dra03} advises tabulating 
the phase function at short wavelengths, rather than employing an analytical expression. 
We have adopted his definition of the phase function, 

\begin{equation}
\label{phasefunc}
\Phi(\theta,\lambda) \equiv 
\frac{1}{\sigma_{\rm sca}(\lambda)} 
\frac{d\sigma_{\rm sca}(\theta,\lambda)}{d\Omega}
\end{equation}

\noindent 
where $\sigma_{\rm sca}$ is the scattering cross section and 
${d\sigma_{\rm sca}(\theta,\lambda)}/{d\Omega}$ is the differential scattering cross section 
per H nucleon, both as a function of wavelength $\lambda$.
The scattering angle, $\theta$, is obtained from 
$\cos(\theta) = [(b/ct)^2 - 1] / [(b/ct)^2 + 1]$ \citep[e.g.,][]{schaefer87}.
The scattering angle is therefore $\theta \approx 9{\fdg}7$ for the SN 2012aw echo.
To compute $\Phi(\theta,\lambda)$ we used the tabulated wavelength-dependent scattering 
properties for $10\arcdeg$ from \citet{dra03} for the \citet{weingartner01} Galactic extinction model 
with $R_V=3.1$.
Furthermore, we also adopted $C_{\rm sca}$ from the \citeauthor{weingartner01}~dust model, with the
update from \citet{dra03}, which also includes the relevant carbonaceous and silicate grain size distributions 
$\phi$.
\citet{vandyk12} argued that the metallicity in the SN environment is consistent with solar metallicity, 
so assuming a Galactic model for the dust in the SN interstellar environment is probably valid.
Finally, we adopted the line-of-sight Galactic foreground visual extinction toward SN 2012aw from
\citet{schlafly11}, i.e., $A_V=0.076$ mag, again, assuming the \citet{cardelli89} reddening law.

\begin{figure}
\figurenum{7}
\epsscale{0.94}
\plotone{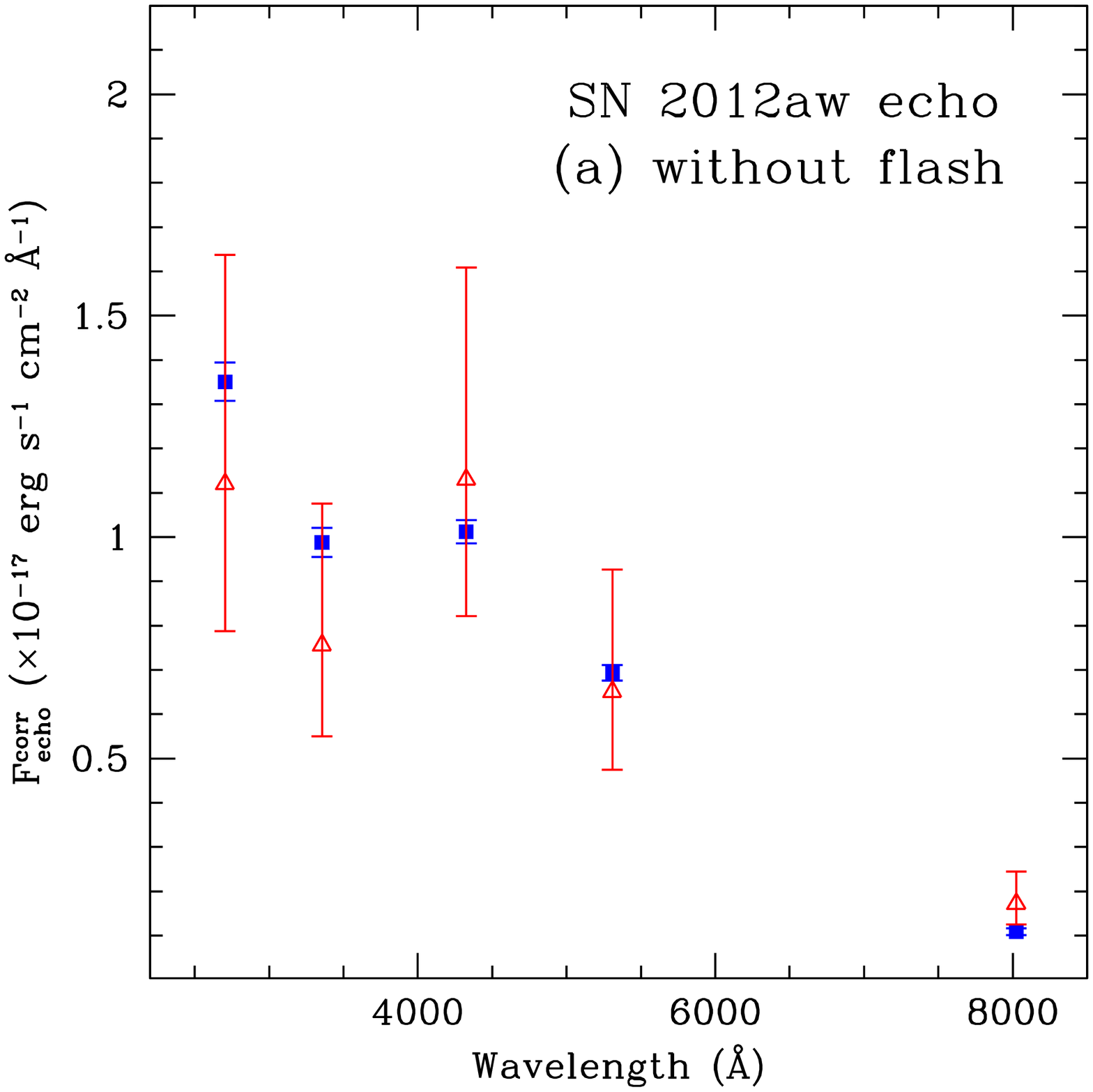}
\plotone{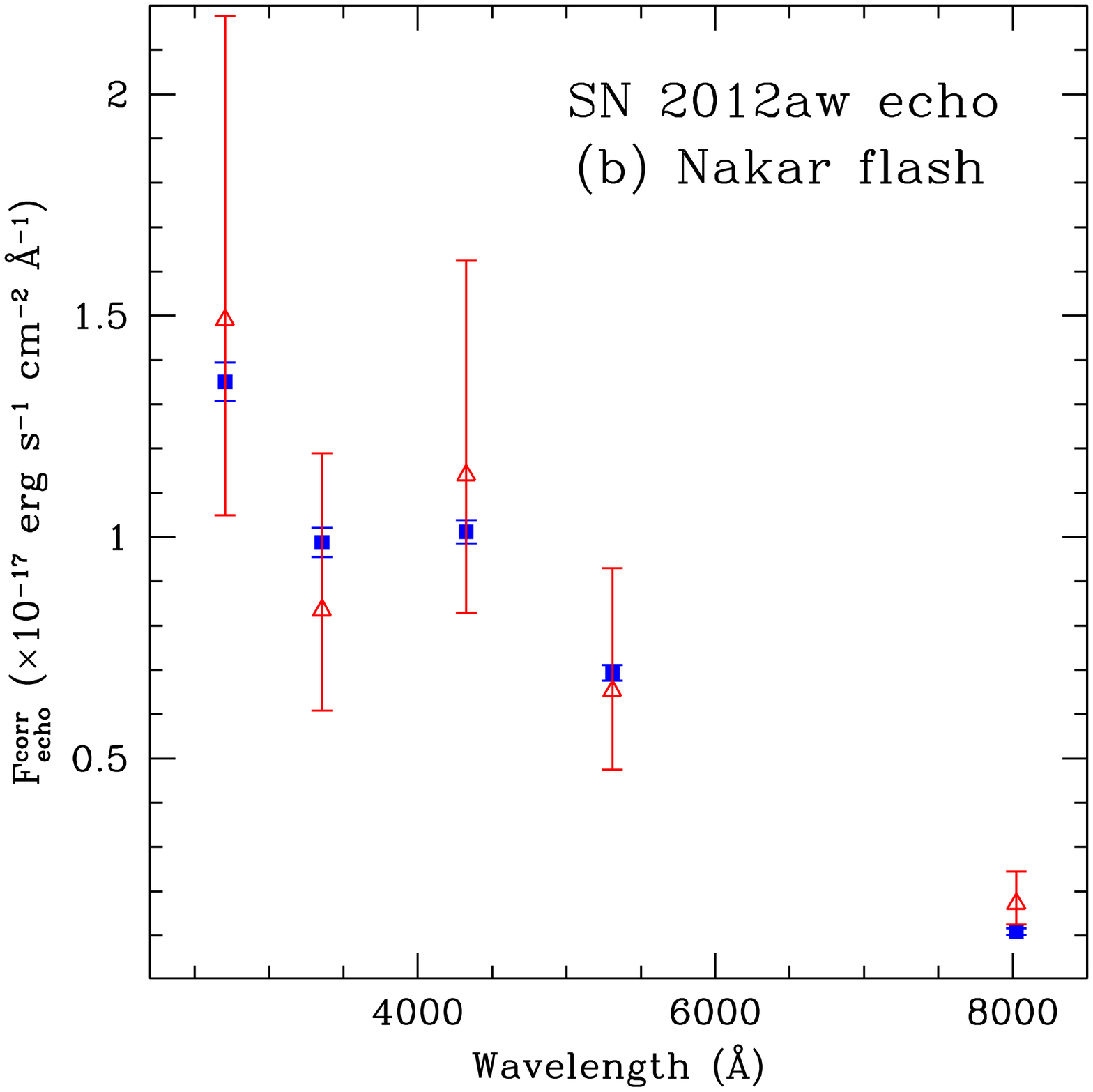}
\plotone{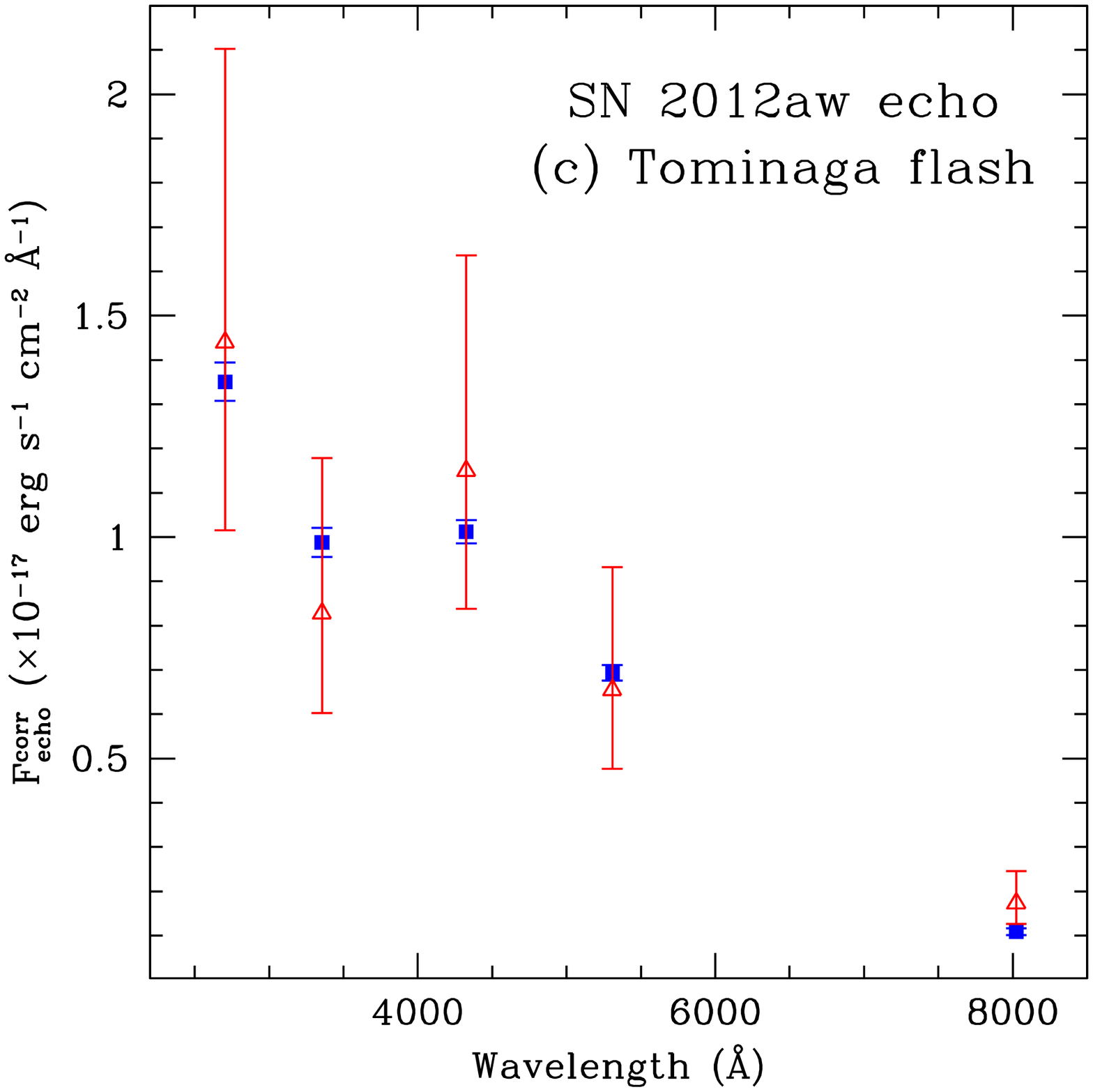}
\caption{The observed light echo flux (solid squares), $F^{\rm corr}_{\rm echo}$, in each of the WFC3 bands, after 
correction for the SN flux, as given in
Table~2. Shown for comparison is the model flux (open triangles) as defined in 
Equations~\ref{echomodel} and \ref{eqresponse}, with an assumed H column density 
$N_H=3.6 \times 10^{20}$ cm$^{-2}$ and reddening by the Galactic foreground 
\citep{schlafly11}, for (a) the integrated observed light curves with no shock-breakout flash; 
(b) the integrated observed light curves with the shock breakout flash model from \citet{nakar10}; and, 
(c) the integrated observed light curves with the shock breakout flash model from \citet{tominaga11}. 
\label{echosed}}
\end{figure}

We show the results of our modeling of the echo in Figure~\ref{echosed}. 
The uncertainty in $F_{\rm model}$ shown in the figure is entirely a result of the uncertainty in $r$, the
distance from the SN to any scattering dust element along the echo.
Additional uncertainties which we have not included can arise from the scattering dust width, 
the inclination of the dust sheet with respect to the line-of-sight
\citep{rest11a,rest12a}, 
and any asymmetries in the SN explosion (\citealt{rest11b,sinnott13}; early-time spectropolarimetry by
\citealt{leonard12} may have indicated large intrinsic polarization and significant asymmetries in the
outer SN ejecta).

The observations constrain $N_H$ in the echo-producing region 
to a narrow range of 3.55--$3.73 \times 10^{20}$ cm$^{-2}$. In Figure~\ref{echosed} we show echo models 
with the average value, $N_H=3.6 \times 10^{20}$ cm$^{-2}$. 
Following the relation between $N_H$ and $A_V$ from \citet{guver09}, this implies that the internal 
$A_V=0.16$ mag for the echo-producing dust.
We would obtain the same estimate for $A_V$ based on the relation from \citet{predehl95}.
We note that \citet{vandyk12} estimated the extinction within the host galaxy, based on the equivalent widths
of the Na~{\sc i} D features in a high-resolution spectrum of the SN near maximum light, to be 
$A_V = 0.17 \pm 0.04$ mag (see also \citealt{bose13}), which is consistent with the implied 
extinction from the echo observations. The SN occurred in a region of M95 seemingly devoid of 
any recent massive star formation, so natal molecular gas and dust may have already dispersed some
time ago. 
We find the close agreement in these estimates of $A_V$ to be sufficiently
reassuring that our light echo models are reasonably correct. 
If we include the Galactic foreground, the total extinction to SN 2012aw is $A_V=0.24$ mag.

The echo model is certainly not perfect: We have only included models for the very early-time light curve 
behavior for the SN and do not know the exact nature of the shock breakout.  We employed interpolated
light curves in our computation of the SN fluences, since the observed photometric coverage in the five 
bands is not entirely complete. We also may have
incorrectly estimated the SN contribution to the total echo light. Possible faint positive residuals can still be 
seen in the SN-subtracted images of the echo. 
Our estimate of the echo radius may well be incorrect. Additionally, our assumed host galaxy distance 
may not be correct; the NASA/IPAC Extragalactic Database (NED) list 21 Cepheid-based distance moduli
which span, to within their uncertainties, a range of about 2.7 Mpc in distance, although the 
uncertainty-weighted mean of these distance modulus estimates, 29.95 mag, is within the uncertainties of
our assumed distance modulus.
If the echo radius were 
actually larger, 
implying $r$ would also be larger, then $N_H$ (and, thus, $A_V$) would correspondingly also have to be
larger; the dust model itself would also change somewhat, as the scattering angle $\theta$ would be 
decreased. (The reverse would be the case if the echo radius were smaller.)
The overall agreement with $A_V$ obtained from the early-time SN spectrum would no longer hold, although
it is conceivable that that estimate is in error.
(\citealt{dallora14} estimated that the host extinction could be as high as $A_V \approx 0.44$--0.59 mag.)
Furthermore, 
the echo-producing dust in M95 may not be as similar to Milky Way diffuse dust as we have assumed.
All of these factors would result in increasing the uncertainty in our overall model.

Nonetheless, we have shown that, to within the uncertainties that we have estimated,  
we can fully and consistently account for the observed flux from 
the echo by assuming that the SN light has been scattered by diffuse interstellar dust. 
The echo light can actually be fit adequately enough without addition of the flash models, although the model
flux in the UV is somewhat low. Including the 
shock breakout flash, using either model, provides a superior representation of the observations in the blue 
and UV. 
We know that a flash must occur, so it is logical to have included these models with the observed light 
curves.
All of the echo models tend to overpredict the flux slightly, relative to the observation, at $I$.

\section{Conclusions}\label{discussion}

We have discovered in the LEGUS multi-wavelength {\sl HST\/} imaging a dust-scattered light echo around 
the SN II-P 2012aw in M95, at $\approx 768$ d after explosion. 
The superior {\sl HST\/} angular resolution was essential; at
$\lesssim 0{\farcs}4$ in diameter, observations from the ground under only truly exceptional conditions 
would have potentially partially resolved the echo.
The echo was quite bright at the time, at
$\sim$21--22 mag in each of the five WFC3/UVIS bands. The echo is a result of the scattering of the total
SN light from interstellar dust in the host galaxy, which we assume to have properties similar to Milky Way
diffuse dust. 
In essence, the light echo is a complete record of the SN's luminosity evolution.
We find from comparing a model of the echo to the observations that the amount of the 
dust extinction in the SN environment responsible for the echo is consistent with the value that was estimated
from observations of the SN itself at early times.
An independent means of constraining $N_H$ might come from X-ray observations of SN 2012aw: 
The SN was observed on 2012 April 11 with the {\sl Chandra X-ray Observatory}, and the data provide only
a modest constraint, with the value of $N_H$ implied by the echo consistent with the limits allowed by the
X-ray spectrum (D.~Pooley, private communication).

The echo around SN 2012aw had become detectable in 2014 as the SN light had become sufficiently dim.
Under the assumption of a homogeneous thin dust sheet in front of the SN, the light echo is always present
and essentially constant in luminosity, even at maximum light,
although the echo luminosity is typically $\sim 10$ mag fainter than the SN at maximum
(see, e.g., \citealt{patat05}). 
If one reasonably assumes that at $\sim 768$ d the total luminosity is dominated by the echo, it is
$\sim 11$ mag less luminous at $V$ than the SN at maximum light, which is roughly in line with 
the prediction for typical interstellar light echoes. From the approximations in 
\citet[][his Section 3]{patat05} 
and adopting our estimates of the SN-sheet distance and the dust optical depth, one 
would obtain a SN-light echo contrast of $\sim 9$ mag. This assumes standard parameters for the dust and 
a homogeneous distribution within a thin dust sheet --- the $\sim 2$ mag difference between the observed
echo brightness and what is predicted could be explained by an inhomogeneous dust distribution, as 
apparently indicated by the asymmetric shape of the echo (see Figure~\ref{figsuper}).
One could conclude from the echo's appearance that there is more dust toward one side of the 
line-of-sight, so that the echo may also have increased in luminosity with time, as a result of increasing
dust optical depth. 

Furthermore, we find that, if our estimate of the SN brightness in the F814W band is correct at
$\approx 23.8$ mag (see Table~2), then the SN at the time was fainter than the 
observed brightness of the SN progenitor star in that band, 23.39 mag \citep{vandyk12,fraser12}. 
We cannot completely rule out that dust in the SN ejecta or immediate SN environment is at least partially 
responsible for the SN's faintness in 2014. However, it was evident that any circumstellar dust in the 
progenitor envelope was destroyed by the X-ray/UV flash \citep{vandyk12}. Adopting a SN expansion 
velocity of 3631 km s$^{-1}$ for SN 2012aw \citep{bose13}, after 2.1 yr the SN would have already swept 
through $\approx 3.4 \times 10^5\ R_{\odot}$, or $\approx 1600$ AU, most probably beyond the immediate 
circumstellar environment. 
Additionally, analysis of {\em Spitzer Space Telescope\/} late-time data (to be presented elsewhere)
shows that SN 2012aw was fading quite rapidly at 3.6 and 4.5 $\mu$m up to day $\sim 868$, when emission
from the SN in these bands was barely detectable. Dust observed at these wavelengths would be relatively 
warm, and we cannot discount that colder SN dust may be present.
Also, relatively small newly-condensed dust masses have been found for SNe II-P
(e.g., \citealt{meikle07,kotak09,meikle11}; although see, e.g., \citealt{matsuura15} in the case of SN 1987A).
A more straightforward conclusion is that the progenitor star has vanished and that 
the RSG identified by \citet{vandyk12} and \citet{fraser12} was almost certainly the progenitor of SN 2012aw.

Finally, the SN 2012aw light echo should be further monitored with {\sl HST}, particularly in the blue, to follow
the evolution of its luminosity, as well as potentially to reveal new or evolving echo structures, and 
better constrain the nature of the echo geometry and of the scattering dust. 
Continued observations will also
place further constraints on the late-time SN brightness and disappearance of the progenitor.
Analysis of any future polarimetric imaging of the echo \citep[e.g.,][]{sparks94,sparks99}, which would allow 
for
an independent distance determination to the host galaxy, however, could be challenging, given the relatively 
compact nature of the SN 2012aw light echo.

\acknowledgments

We are grateful for the comments provided by the expert referee, which significantly improved this paper.
We appreciate helpful discussions we had with Luc Dessart and the generosities of both Ehud Nakar
and Nozomu Tominaga. We also appreciate David Pooley sharing his analysis of the {\sl Chandra\/} data
on SN 2012aw in advance of publication.
Based on observations made with the NASA/ESA {\sl Hubble Space Telescope}, obtained at the Space 
Telescope Science Institute, which is operated by the Association of Universities for Research in Astronomy, 
Inc., under NASA contract NAS 5-26555. These observations are associated with program \#13364.
Support for this program was provided by NASA through a grant from the Space Telescope Science Institute.
This research made use of the NASA/IPAC Extragalactic Database (NED), which is operated by the Jet 
Propulsion Laboratory, California Institute of Technology, under contract with NASA. 
AH acknowledges funding by Spanish MINECO under grants AYA2012-39364-C02-01 and SEV 2011-0187-01.
SEdM acknowledges support by a Marie Sklodowska-Curie Reintegration Fellowship (H2020-MSCA-IF-2014, project 
id 661502) awarded by the European Commission.


\begin{thebibliography}{}
\bibitem[Bayless et al.(2013)]{bayless13} Bayless, A.~J., 
Pritchard, T.~A., Roming, P.~W.~A., et al.\ 2013, \apjl, 764, L13
\bibitem[Bond et al.(1990)]{bond90} Bond, H.~E., Gilmozzi, R., Meakes, M.~G., et al.\ 1990, 
\apj, 354, L49
\bibitem[Bond et al.(2003)]{bond03} Bond, H.~E., Henden, A., Levay, Z.~G., et al.\ 2003, \nat, 
422, 405 
\bibitem[Bose et al.(2013)]{bose13} Bose, S., Kumar, B., Sutaria, F., et al.\ 2013, \mnras, 433, 1871
\bibitem[Calzetti et al.(2015)]{calzetti15} Calzetti, D., Lee, 
J.~C., Sabbi, E., et al.\ 2015, \aj, 149, 51
\bibitem[Cappellaro et al.(2001)]{cappellaro01} Cappellaro, E.,
Patat, F., Mazzali, P.~A., et al.\ 2001, \apjl, 549, L215
\bibitem[Cardelli, Clayton, \& Mathis(1989)]{cardelli89} Cardelli, J. A.,
Clayton, G. C., \& Mathis, J. S. 1989, \apj, 345, 245
\bibitem[Chevalier(1986)]{che86} Chevalier, R.~A.\ 1986, \apj, 308, 225
\bibitem[Couderc(1939)]{couderc39} Couderc, P.\ 1939, Annales d'Astrophysique, 2, 271
\bibitem[Crotts(1988)]{crotts88} Crotts, A. P. S. 1988, \apj, 333, L51 
\bibitem[Crotts(2014)]{crotts14} Crotts, A.\ 2014, \apjl, submitted (arXiv:1409.8671)
\bibitem[Crotts \& Kunkel(1991)]{crotts91} Crotts, A.~P.~S., \& Kunkel, W.~E.\ 1991, \apjl, 366, L73
\bibitem[Crotts et al.(1992)]{crotts92} Crotts, A.~P.~S., Landsman, W.~B., Bohlin, R.~C., et al.\ 1992, 
\apjl, 395, L25
\bibitem[Dall'Ora et al.(2014)]{dallora14} Dall'Ora, M., Botticella, M.~T., Pumo, M.~L., et al.\ 2014, \apj, 787, 
139
\bibitem[Dolphin(2000)]{dolphin00} Dolphin, A.~E. 2000, PASP, 112, 1383
\bibitem[Draine(2003)]{dra03} Draine, B.~T. 2003, \apj, 598, 1017
\bibitem[Drozdov et al.(2014)]{drozdov14} Drozdov, D., Leising, M.~D., Milne, P.~A., et al.\ 2014, 
arXiv:1410.8190 
\bibitem[Ekstr{\"o}m et al.(2012)]{ekstrom12} Ekstr{\"o}m, S., Georgy, C., Eggenberger, P., et al.\ 2012, \aap, 
537, A146
\bibitem[Eldridge et al.(2008)]{eldridge08} Eldridge, J.~J., 
Izzard, R.~G., \& Tout, C.~A.\ 2008, \mnras, 384, 1109
\bibitem[Fraser et al.(2012)]{fraser12} Fraser, M., Maund, J.~R., Smartt, S.~J., et al.\ 2012, \apjl, 759, L13
\bibitem[Freedman et al.(2001)]{freedman01} Freedman, W.~L., 
Madore, B.~F., Gibson, B.~K., et al.\ 2001, \apj, 553, 47
\bibitem[Fruchter et al.(2010)]{fruchter10} Fruchter, A.~S., et al.\ 2010, 2010 Space Telescope Science 
Institute Calibration Workshop, p.~382-387
\bibitem[Gezari et al.(2008)]{gezari08} Gezari, S., Dessart, L., Basa, S., et al.\ 2008, \apjl, 683, L131
\bibitem[Gezari et al.(2015)]{gezari15} Gezari, S., Jones, D.~O., Sanders, N.~E., et al.\ 2015, \apj, in press 
(arXiv:1502.06964)
\bibitem[Gonzaga et al.(2012)]{gonzaga12} Gonzaga, S., Hack, W., Fruchter, A., \& Mack, J. (eds.) 2012, The 
DrizzlePac Handbook (Baltimore: STScI)
\bibitem[Gouiffes et al.(1988)]{gouiffes88} Gouiffes, C., Rosa, M., Melnick, J., et al.\ 1988, \aap, 198, L9
\bibitem[G{\"u}ver \& {\"O}zel(2009)]{guver09} G{\"u}ver, T., \& {\"O}zel, F.\ 2009, \mnras, 400, 2050
\bibitem[Hamuy et al.(2001)]{hamuy01} Hamuy, M., Pinto, P.~A., Maza, J., et al.\ 2001, \apj, 558, 615
\bibitem[Henyey \& Greenstein(1941)]{hen41} Henyey, L.~C., \& Greenstein, J.~L. 1941, \apj, 93, 70
\bibitem[Jerkstrand et al.(2014)]{jerkstrand14} Jerkstrand, A., Smartt, S.~J., Fraser, M., et al.\ 2014, \mnras, 
439, 3694
\bibitem[Kapteyn(1901)]{kapteyn01} Kapteyn, J.~C.\ 1901, Astronomische Nachrichten, 157, 201
\bibitem[Kochanek et al.(2012)]{kochanek12} Kochanek, C.~S., Khan, R., \& Dai, X.\ 2012, \apj, 759, 20
\bibitem[Kotak et al.(2009)]{kotak09} Kotak, R., Meikle, W.~P.~S., Farrah, D., et al.\ 2009, \apj, 704, 306
\bibitem[Krause et al.(2005)]{krause05} Krause, O., Rieke, G.~H., Birkmann, S.~M., et al.\ 2005, Science, 
308, 1604 
\bibitem[Krause et al.(2008)]{krause08} Krause, O., Tanaka, M., Usuda, T., et al.\ 2008, \nat, 456, 617
\bibitem[Leonard et al.(2002)]{leonard02} Leonard, D.~C., Filippenko, A.~V., Gates, E.~L., et al.\ 2002, \pasp, 114, 35
\bibitem[Leonard et al.(2012)]{leonard12} Leonard, D.~C., 
Pignata, G., Dessart, L., et al.\ 2012, The Astronomer's Telegram, 4033, 1
\bibitem[Liu et al.(2003)]{liu03} Liu, J.-F., Bregman, J.~N., \& Seitzer, P.\ 2003, \apj, 582, 919
\bibitem[Matsuura et al.(2015)]{matsuura15} Matsuura, M., Dwek, 
E., Barlow, M.~J., et al.\ 2015, \apj, 800, 50
\bibitem[Mattila et al.(2008)]{mattila08} Mattila, S., Meikle, 
W.~P.~S., Lundqvist, P., et al.\ 2008, \mnras, 389, 141
\bibitem[Meikle et al.(2006)]{meikle06} Meikle, W.~P.~S., 
Mattila, S., Gerardy, C.~L., et al.\ 2006, \apj, 649, 332
\bibitem[Meikle et al.(2007)]{meikle07} Meikle, W.~P.~S., 
Mattila, S., Pastorello, A., et al.\ 2007, \apj, 665, 608
\bibitem[Meikle et al.(2011)]{meikle11} Meikle, W.~P.~S., Kotak, 
R., Farrah, D., et al.\ 2011, \apj, 732, 109
\bibitem[Miller et al.(2010)]{miller10} Miller, A.~A., Smith, 
N., Li, W., et al.\ 2010, \aj, 139, 2218
\bibitem[Nakar \& Sari(2010)]{nakar10} Nakar, E., \& Sari, R.\ 2010, \apj, 725, 904 
\bibitem[Patat(2005)]{patat05} Patat, F. 2005, \mnras, 357, 1161
\bibitem[Predehl \& Schmitt(1995)]{predehl95} Predehl, P., \& Schmitt, J.~H.~M.~M.\ 1995, \aap, 293, 889
\bibitem[Prieto et al.(2014)]{prieto14} Prieto, J.~L., Rest, A., Bianco, F.~B., et al.\ 2014, \apjl, 787, LL8
\bibitem[Quimby et al.(2007)]{quimby07} Quimby, R.~M., Wheeler, 
J.~C., H{\"o}flich, P., et al.\ 2007, \apj, 666, 1093
\bibitem[Quinn et al.(2006)]{quinn06} Quinn, J.~L., Garnavich, P.~M., Li, W., et al.\ 2006, \apj, 652, 512
\bibitem[Rest et al.(2005)]{rest05} Rest, A., Suntzeff, N.~B., Olsen, K., et al.\ 2005, \nat, 438, 1132 
\bibitem[Rest et al.(2008a)]{rest08a} Rest, A., Matheson, T., Blondin, S., et al.\ 2008a, 
\apj, 680, 1137 
\bibitem[Rest et al.(2008b)]{rest08b} Rest, A., Welch, D.~L., Suntzeff, N.~B., et al.\ 2008b, 
\apjl, 681, L81
\bibitem[Rest et al.(2011a)]{rest11a} Rest, A., Sinnott, B., Welch, D.~L., et al.\ 2011a, \apj, 732, 2
\bibitem[Rest et al.(2011b)]{rest11b} Rest, A., Foley, R.~J., Sinnott, B., et al.\ 2011b, \apj, 732, 3
\bibitem[Rest et al.(2012a)]{rest12a} Rest, A., Prieto, J.~L., Walborn, N.~R., et al.\ 2012a, \nat, 482, 375 
\bibitem[Rest et al.(2012b)]{rest12b} Rest, A., Sinnott, B., \& Welch, D.~L.\ 2012b, PASA, 29, 466 
\bibitem[Ritchey(1901)]{ritchey01} Ritchey, G.~W.\ 1901, \apj, 14, 293
\bibitem[Sana et al.(2012)]{sana12} Sana, H., de Mink, S.~E., de Koter, A., et al.\ 2012, Science, 337, 444
\bibitem[Schaefer(1987)]{schaefer87} Schaefer, B.~E.\ 1987, \apjl, 323, L47
\bibitem[Schawinski et al.(2008)]{schawinski08} Schawinski, K., Justham, S., Wolf, C., et al.\ 2008, 
Science, 321, 223
\bibitem[Schlafly \& Finkbeiner(2011)]{schlafly11} Schlafly, E.~F., \& Finkbeiner, D.~P.\ 2011, \apj, 737, 103
\bibitem[Sinnott et al.(2013)]{sinnott13} Sinnott, B., Welch,
D.~L., Rest, A., Sutherland, P.~G., \& Bergmann, M.\ 2013, \apj, 767, 45
\bibitem[Sokoloski et al.(2013)]{sokoloski13} Sokoloski, J.~L., Crotts, A.~P.~S., Lawrence, S., \& 
Uthas, H.\ 2013, \apjl, 770, L33
\bibitem[Sparks(1994)]{sparks94} Sparks, W. B. 1994, \apj, 433, 19 
\bibitem[Sparks(1996)]{sparks96} Sparks, W. B. 1996, \apj, 470, 195 
\bibitem[Sparks et al.(1999)]{sparks99} Sparks, W.~B., Macchetto, F., Panagia, N., et al.\ 1999, \apj, 523, 585
\bibitem[Sugerman(2003)]{sugerman03} Sugerman, B.~E.~K. 2003, \aj, 126, 1939 
\bibitem[Sugerman(2005)]{sugerman05} Sugerman, B.~E.~K.\ 2005, \apjl, 632, L17
\bibitem[Sugerman et al.(2012)]{sugerman12} Sugerman, B.~E.~K., 
Andrews, J.~E., Barlow, M.~J., et al.\ 2012, \apj, 749, 170
\bibitem[Sugerman \& Crotts(2002)]{sugerman02} Sugerman, B.~E.~K., \& Crotts,
A.~P.~S. 2002, \apj, 581, L97
\bibitem[Tamburro et al.(2008)]{tamburro08} Tamburro, D., Rix, H.-W., Walter, F., et al.\ 2008, \aj, 136, 2872
\bibitem[Tominaga et al.(2011)]{tominaga11} Tominaga, N., Morokuma, T., Blinnikov, S.~I., et al.\ 2011, \apjs, 193, 20
\bibitem[Van Dyk(2013)]{vandyk13} Van Dyk, S.~D.\ 2013, \aj, 146, 24
\bibitem[Van Dyk et al.(2006)]{vandyk06} Van Dyk, S.~D., Li, W., \& Filippenko, A.~V.\ 2006, \pasp, 118, 351 
\bibitem[Van Dyk et al.(2012)]{vandyk12} Van Dyk, S.~D., Cenko, S.~B., Poznanski, D., et al.\ 2012, \apj, 
756, 131
\bibitem[Wang et al.(2008)]{wang08} Wang, X., Li, W., Filippenko, A.~V., et al.\ 2008, \apj, 677, 1060
\bibitem[Weingartner \& Draine(2001)]{weingartner01} Weingartner, J.~C., \&
Draine, B.~T. 2001, \apj, 548, 29
\bibitem[Welch et al.(2007)]{welch07} Welch, D.~L., Clayton, 
G.~C., Campbell, A., et al.\ 2007, \apj, 669, 525
\bibitem[Xu et al.(1994)]{xu94} Xu, J., Crotts, A.~P.~S., \& Kunkel, W.~E.\ 1994, \apj, 435, 274
\bibitem[Xu et al.(1995)]{xu95} Xu, J., Crotts, A.~P.~S., \& Kunkel, W.~E.\ 1995, \apj, 451, 806 (Erratum: 
1996, \apj, 463, 391)
\bibitem[Yadav et al.(2014)]{yadav14} Yadav, N., Ray, A., Chakraborti, S., et al.\ 2014, \apj, 782, 30 
\end{thebibliography}
\end{document}